\documentclass[preprint,superscriptaddress,floatfix,preprintnumbers,nofootinbib,altaffilletter]{revtex4-1}
\pdfoutput=1
\usepackage{graphicx}  
\usepackage{dcolumn}   
\usepackage{bm}        
\usepackage{amssymb}   
\usepackage[utf8]{inputenc}
\usepackage[OT1]{fontenc}
\usepackage{yfonts,amsmath,amsthm,amsfonts,amssymb,amscd}
\usepackage{enumerate}
\usepackage{fancyhdr}
\usepackage{mathrsfs}
\usepackage{cancel}
\usepackage{slashed}
\usepackage[flushleft]{threeparttable}
\usepackage[colorlinks=true, citecolor=purple, linkcolor=blue]{hyperref}
\usepackage{url}
\usepackage{makecell,booktabs}
\usepackage{braket}
\usepackage{relsize}
\usepackage{tikz}

\begin{document}

\preprint{EFI-21-2}

\title{Challenges  for a QCD Axion at the 10 MeV Scale}
\author{Jia Liu}
\email{jialiu@pku.edu.cn}
\affiliation{School of Physics and State Key Laboratory of Nuclear Physics and Technology, Peking University, Beijing 100871, China}
\affiliation{Center for High Energy Physics, Peking University, Beijing 100871, China}

\author{Navin McGinnis}
\email{nmcginnis@triumf.ca}
\affiliation{TRIUMF, 4004 Westbrook Mall, Vancouver, BC, Canada, V6T 2A3}
\affiliation{High Energy Physics Division, Argonne National Laboratory, Argonne, IL 60439, USA}

\author{Carlos E.M. Wagner}
\email{cwagner@uchicago.edu}
\affiliation{High Energy Physics Division, Argonne National Laboratory, Argonne, IL 60439, USA}
\affiliation{\mbox{Physics Department and Enrico Fermi Institute, University of Chicago, Chicago, IL 60637}}
\affiliation{\mbox{Kavli Institute for Cosmological Physics, University of Chicago, Chicago, IL 60637, USA}}

\author{Xiao-Ping Wang}
\email{hcwangxiaoping@buaa.edu.cn}
\affiliation{School of Physics, Beihang University, Beijing 100083, China}
\affiliation{Beijing Key Laboratory of Advanced Nuclear Materials and Physics, Beihang University, Beijing 100191, China}

\begin{abstract}
We report on an interesting realization of the QCD axion, with mass in the range $\mathcal{O}(10)$ MeV. 
It has previously been shown that although this scenario is stringently constrained from multiple sources, 
 the model remains viable for a range of parameters that  leads to an explanation of the Atomki experiment anomaly.  
 In this article we study in more detail the additional
constraints proceeding from recent low energy experiments and study the compatibility of the allowed parameter
space with the one leading to consistency of the most recent measurements of the electron anomalous magnetic
moment and the fine structure constant.  We further  provide an ultraviolet completion of this axion variant and show 
the conditions under which it may lead to the observed quark masses and CKM mixing angles, and remain consistent
with experimental constraints on the extended scalar sector appearing in this Standard Model extension. In particular,
the decay of the Standard Model-like Higgs boson into two light axions may be relevant and leads to a novel Higgs boson 
signature that may be searched for at the LHC in the near future. 
\end{abstract}

\maketitle
\tableofcontents

\section{Introduction}

The Standard Model (SM) provides a consistent description of particle interactions, based on a local and
renormalizable gauge theory~\cite{Zyla:2020zbs}.  The gauge interactions of the theory have been tested with great
precision, while the Yukawa interactions of quarks and leptons with the recently discovered Higgs boson are being
currently tested at the Large Hadron collider (LHC).  In the SM, these Yukawa
interactions lead to a rich flavor structure based on the diversity of quark and lepton masses
and the CKM and PMNS matrices in the quark and lepton sector, respectively.   These matrices
include CP violating phases that are sufficient to explain all CP-violating phenomena observed
in B and K meson, as well as neutrino experiments.  

Although the gluon interactions with quarks preserve the  CP symmetry, there is a CP violating
interaction in the SM QCD sector that is related to the interactions induced by  the $\theta_{\rm QCD}$ 
parameter.  Even in the absence of a tree-level value, a non-vanishing value of $\theta_{\rm QCD}$ 
would be generated via the chiral anomaly, by the chiral phase redefinition associated with making the 
quark masses real.  In general, hence, the physical value of $\theta_{\rm QCD}$ comes from the
addition of the tree-level value and the phase of the determinant of the quark masses, and is
naturally of order one.  In the presence of an unsuppressed  value of $\theta_{\rm QCD}$,  however, 
the neutron electric dipole moment (EDM) would acquire an unobserved sizable value~\cite{Crewther:1979pi}. The experiment 
measuring neutron EDM has found that such CP phase should be smaller than $10^{-10}$ \cite{Abel:2020gbr}, 
which leads to a severe fine tuning problem in the SM~\cite{Zyla:2020zbs}.

One elegant way to solve the neutron EDM problem is to introduce a chiral symmetry, the so-called PQ symmetry, which is spontaneously broken, leading to a 
Goldstone boson, $a$, which has been called the axion~\cite{Peccei:1977hh, Peccei:1977ur, Weinberg:1977ma, Wilczek:1977pj}.  The axion interactions lead to 
a field dependent $\theta_{\rm QCD}$ angle, and the strong interactions induce a non-trivial potential which is minimized at vanishing values of $\theta_{\rm QCD}$,
therefore leading to the disappearance of  the CP interactions in the strong sector in the physical vacuum and the solution of the strong CP problem.
 
Being associated with a Goldstone boson, the axion field has an approximate shift symmetry, thus it can be naturally light. Its couplings to other particles may be rendered 
small due to the suppression from large decay constant, $f_a$, proportional to the vacuum expectation value of the field that breaks the chiral PQ symmetry.  Since the 
$f_a$ scale is not necessarily related to the weak or QCD scales, the axion mass and couplings to fermions and gauge bosons, which are proportional to $1/f_a$, 
can span many orders of magnitude.  Thus, there is a rich phenomenology to explore this wide region of parameter 
space with different experimental setups~\cite{Blumlein:1990ay, Blumlein:1991xh, Kim:2008hd, Kawasaki:2013ae, Graham:2015ouw, Zyla:2020zbs}. 
The original \textit{visible} axion scenario \cite{Weinberg:1977ma, Wilczek:1977pj}, where
$f_a$ is related to the weak scale, has been severely constrained by laboratory experiments \cite{Blumlein:1990ay, Blumlein:1991xh, Graham:2015ouw, Zyla:2020zbs}. There are also \textit{invisible} axion scenarios, 
with $f_a$ much larger than electroweak scale, where the axion remains hidden due to its ultralight mass and ultra small couplings.  A realization of these invisible axions
are given in the so-called KSVZ  and DFSZ axion scenarios~\cite{Kim:1979if, Shifman:1979if,Dine:1981rt, Zhitnitsky:1980tq}.

Regarding the \textit{visible} axion, there are efforts \cite{Krauss:1987ud, Alves:2017avw, Alves:2020xhf} to save it from experimental constraints by making it  \textit{pion-phobic}.
In this scenario, the mixing with the neutral pion is suppressed by choosing appropriate PQ charges and an accidental cancellation arising from the low energy quark mass and coupling parameters. This \textit{visible} axion scenario has been combined with the recent experimental anomalies from Atomki collaboration \cite{Krasznahorkay:2015iga, Krasznahorkay:2019lyl}.  The experiment looks for high excitation states of nuclei from ${}^8$Be and ${}^4$He, which de-excite into ground states and emit
a pair of electron and positron. A peak was observed in the angular correlation of $e^- e^+$, which can be interpreted as an intermediate pseudo-scalar particle with mass around $17$~MeV produced in the de-excitation and later decay into $e^- e^+$~\cite{Ellwanger:2016wfe, Alves:2017avw, DelleRose:2018pgm, Alves:2020xhf, Feng:2020mbt}.
This model may be tested with multi-lepton signatures when light mesons decay to multiple axion states~\cite{Hostert:2020xku}.
It may also lead to significant dark matter electronic scattering cross-section in the dark matter direct detection~\cite{Buttazzo:2020vfs}.
Recently, it has been suggested that  higher order SM effects may lead to an effect similar to the one observed by the Atomki experiment~\cite{Aleksejevs:2021zjw}. However, further theoretical and experimental analyses are needed to determine whether or not the observed peak may indeed be accounted by SM effects.  
In this work we shall assume that this is not the case and identify the associated pseudo-scalar particle with a 17 MeV~QCD axion.  We will, however, comment on
how the constraints on the proposed QCD axion at the 10~MeV scale would be modified if its resonant signal  were absent. In general, since $m_a \simeq m_\pi f_\pi/f_a$, an axion mass in the 10~MeV range implies $f_a \simeq 1$~GeV.  For this scenario to survive the current experimental constraints, the axion must couple only to first generation quarks and charged leptons with suitable couplings determined by their masses, their QCD charges and $f_a$~\cite{Alves:2017avw, Alves:2020xhf}.

The recent accurate measurement of the fine structure constant~\cite{Parker:2018vye} revealed that there is a mild discrepancy with the electron magnetic dipole moment measurements \cite{Hanneke:2008tm, Hanneke:2010au}, compared to the SM theoretical calculation \cite{Aoyama:2014sxa, Mohr:2015ccw}.   The discrepancy, $\Delta a_e$, associated with this measurement used to be negative and had a significance of about $2.4~\sigma$. On the other hand, a very recent, more accurate measurement of the fine structure constant~\cite{Morel:2020dww} suggests a smaller discrepancy for $\Delta a_e$ and with positive sign.
Since the axion has to decay to electron-positron pairs promptly in the Atomki experiment, it has a sizable coupling to electron. As a pseudo-scalar, it can naturally give a negative contribution to electron $(g-2)_e$
at the one-loop level~\cite{Liu:2018xkx}.  Although positive correction would be therefore
in tension with the one-loop corrections induced by the 17 MeV axion particle, there are relevant two loop corrections that, depending on the axion-$\eta$ meson mixing, may lead to a positive value of~$\Delta a_e$.

In addition to experimental constraints from precision measurement of mesons, this scenario also is constrained by low energy electron collider and beam dump experiments. 
Therefore, it is interesting to ask if this 17 MeV visible axion can evade all current experimental constraints and also be consistent with the current measurements of  $(g-2)_e$ and the fine structure constant.  
It is also important to find a field theoretical realization of this axion. The most natural realizations are associated  with new physics at the weak scale and are therefore subject to further constraints beyond those associated with the effective low energy axion model.  In this article, we will present an extension of the SM which leads to a 10~MeV axion and study the associated constraints on this model. 

This article is organized as follows. In section~\ref{sec:effmodel}, 
we review the properties of this visible axion model and how the most relevant experimental constraints may be avoided under the \textit{pion-phobic} limit.
In section~\ref{sec:electroncoupling}, we consider the electron and photon couplings to the axion and show the region of parameters which can lead to a consistency between the measured values of $(g-2)_e$
and $\alpha$, what leads to a relation between the axion electron and photon couplngs. In section~\ref{sec:Atomki}, we further show how it can survive the constraints coming from low energy electron
colliders and beam dump experiments, while at the same time lead to an explanation of the Atomki experimntal constraints. Furthermore, in section~\ref{sec:model},
we construct a concrete UV model and show how this scenario can couple to only the first generation fermions while successfully generating the CKM matrix.  We also 
discuss the additional constraints implied by the new physics associated with this scenario. We reserve section~\ref{sec:conclusions}
for our conclusions.

\section{Effective Low Energy Model}
\label{sec:effmodel}
We consider the effective theory for a variant of the QCD axion which couples mostly to the first generation fermions in the SM. While the low-energy structure of this setup has appeared some time ago~\cite{Bardeen:1986yb, Krauss:1987ud}, it has recently been discussed how this scenario remains robust to a variety of constraints~\cite{Alves:2017avw,Alves:2020xhf}. From the ultraviolet (UV) perspective, for the axion to couple only to $u$ and $d$ quarks in the infrared (IR), the PQ breaking mechanism must allow for the generation of the appropriate quark and lepton masses as well as the  CKM matrix. 
 We discuss how this is achieved in a particular UV model later in section~\ref{sec:model}. 

From the IR perspective, the relevant low-energy interactions are simply written as

\begin{equation}
\mathcal{L}_{int}= \sum_{f=e,u,d,}m_{f}e^{iQ_{f}a/f_{a}}\bar{f}_{L}f_{R}  + h.c.,
\label{eq:lag}
\end{equation}
where $Q_{f}$ denotes the PQ charge of the fermion $f$, and $f_{a}$ is the axion decay  constant, which, as emphasized in the introduction must be $f_a \simeq 1$~GeV. For the following discussion, it will be convenient to work with the linear axion couplings to fermions. For instance, the coupling to first generation quark and leptons in this basis is simply given by

\begin{equation}
\sum_f g_a^f i a \bar{f}\gamma_5 f, \quad g^{f}_{a} = \frac{Q_{f}m_{f}}{f_{a}}.
\label{eq:fermioncoup}
\end{equation}

At energies below the QCD scale, the couplings of the axion to $u$ and $d$ quarks will induce mixing of the axion in particular to the pion as well as other pseudo-scalar mesons. In this regime, the physical axion state and meson mixing angles can be determined by standard chiral perturbation theory ($\chi$PT) techniques. For detailed explanations and results of this procedure we refer to Refs.~\cite{Alves:2017avw,Alves:2020xhf}. The same mixing angles can be used to parameterize the effective operator controlling the axion coupling to photons

\begin{equation}
g^{\gamma\gamma}_{a}=\frac{\alpha}{4\pi f_{\pi}}\left(\theta_{a\pi} + \frac{5}{3}\theta_{a\eta_{ud}} + \frac{\sqrt{2}}{3}\theta_{a\eta_{s}}\right),
\label{eq:diphoton}
\end{equation}
where $\theta_{a\pi}$ is the mixing angle between the axion and neutral pion, and $\theta_{a\eta_{ud,s}}$ are the mixing angles with the eta-mesons in the light-heavy basis. The relationships between these parameters and other bases for the meson octect in leading order of $\chi$PT are given also in Ref.~\cite{Alves:2017avw}.  For the purpose of this work, it is important to note that for $Q_{u},Q_{d}\simeq\mathcal{O}(1)$, the natural ranges for $\theta_{a\eta_{ud}}$ and $\theta_{a\eta_{s}}$ are in the range of $\mathcal{O}(10^{-3} - 10^{-2})$~\cite{Alves:2017avw}.

Searches for axions in the decay products of pions presents the most stringent constraints for the setup we consider. In particular, the bound on the charged pion decay $BR(\pi^{+}\rightarrow e^{+}\nu_{e}(a\rightarrow e^{-}e^{+}))<0.5\times 10^{-10}$ is quite severe~\cite{Eichler:1986nj}. For $m_{a}=\mathcal{O}(10)$ MeV, this translates to a bound on the mixing angle as  

\begin{equation}
|\theta_{a\pi}|\lesssim \frac{1\times 10^{-4}}{\sqrt{{\rm BR}(a\rightarrow e^{+}e^{-})}} .
\label{eq:pion_mixing_bound}
\end{equation}
As we will see the range of couplings needed for $(g-2)_{e}$ will require the axion width into photons to be subdominant with respect to the one into electrons.  Hence, if decays only to SM particles 
are considered, one expects ${\rm BR}(a\rightarrow e^{+}e^{-}) \simeq 1$. 
Thus, this variant of the axion is \textit{pion-phobic}~\cite{Alves:2017avw,Alves:2020xhf}, in the sense that the mixing of the axion with the pion is below its natural values.  Let us emphasize that, in leading order of perturbation theory
\begin{equation}
m_a^2  \simeq (Q_{u} + Q_{d})^2 \frac{m_u m_d}{(m_u + m_d)^2} \frac{m_\pi^2 f_\pi^2}{f_a^2}
\end{equation}
and 
\begin{equation}
\theta_{a\pi} \simeq \frac{(Q_{d} m_d - Q_{u}m_u)}{(m_u + m_d)} \frac{f_\pi}{f_a},
\end{equation}
where we have chosen an arbitrary normalization of $f_{a}$ with respect to the PQ charges. Since the ratio $m_u/m_d \simeq 0.47 \pm 0.06$~\cite{Zyla:2020zbs}, a cancellation of $\theta_{a \pi} \lesssim 10^{-4}$ demands $Q_{u} \simeq 2 Q_{d}$.  Even assuming 
this relation between the quark PQ charges, the value of the mixing $\theta_{a\pi}$ becomes naturally of the order of a few times $10^{-3}$ and hence an accidental cancellation between the leading
order contribution associated with the precise value of the up and down quark masses and the higher order contributions must occur for this scenario to be viable.  

One could in principle alleviate this aspect of the model assuming that the axion width is dominated by invisible decays, allowing the reduction of the axion decay branching ratio into electrons. 
This possibility, however, is strongly constrained by charged Kaon decays constraints, in particular the decay $K^+ \to (\pi^+  + $ Missing Energy). Indeed, considering a neutral pion state with
a relevant mixing with an axion with dominant invisible decays leads to a bound on the axion-pion mixing $\theta_{a\pi} \lesssim 10^{-4}$, and therefore this scenario does not lead to a relaxation
of the pion axion mixing bound. 

One interesting aspect of a mixing angle $\theta_{a\pi}$ below but close to  $10^{-4}$ is that it leads to a possible resolution of the so-called KTeV anomaly, related to a  decay branching ratio of the
neutral pion into pairs of electrons larger than the one expected in the SM~\cite{Alves:2017avw,Alves:2020xhf}. For this reason, for the rest of this article, we will fix the pion mixing to 
values close to the upper bound, $\theta_{a\pi} = 0.7 \times 10^{-4}$. Apart from the KTeV anomaly,
however, the phenomenological properties of this model  do not change by setting this mixing angle to any other value consistent with the Kaon decay bounds.

\section{$(g-2)_{e}$ }
\label{sec:electroncoupling}

\begin{figure}[htb]
	\centering
	\includegraphics[scale=0.5]{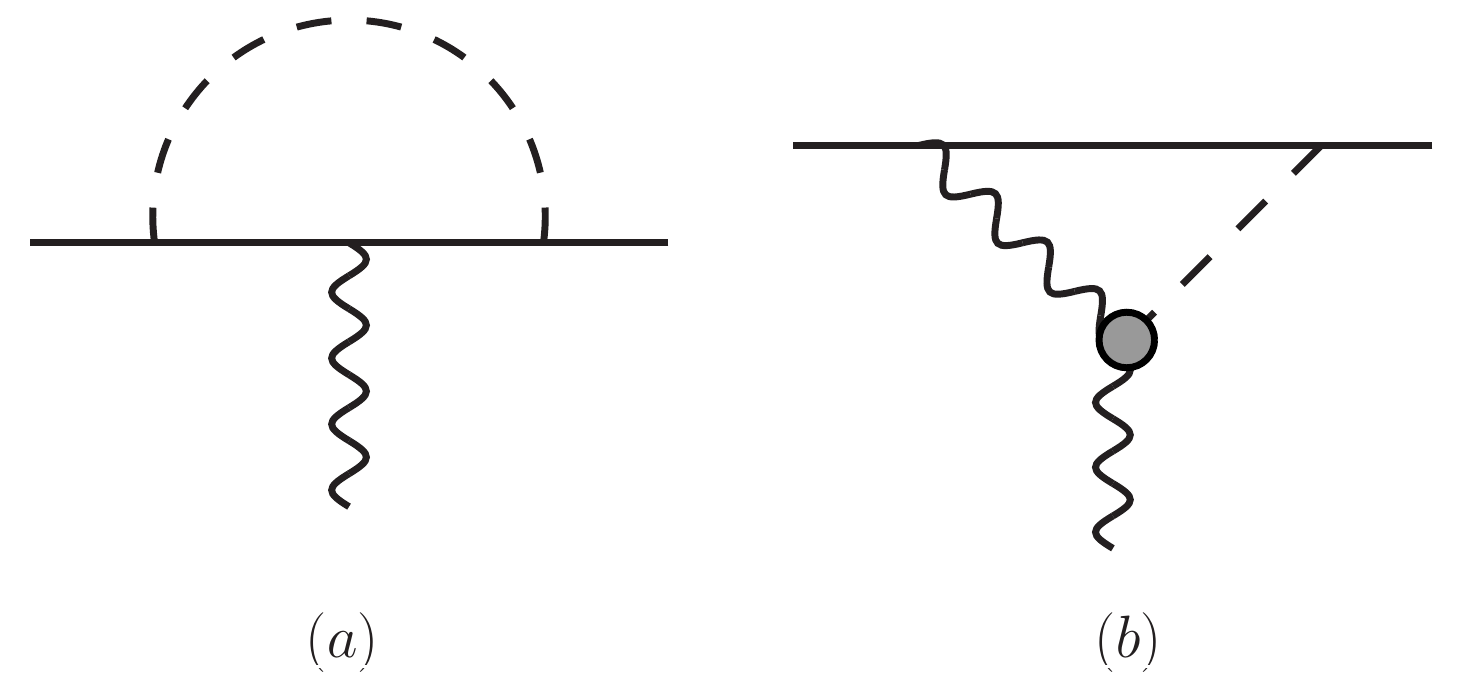}
	\caption{1- and 2-loop contributions to $(g-2)_{e}$ from a pseudo-scalar with couplings to the first generation fermions. The effective vertex is generated at low energies through the meson mixing angles, Eq. (\ref{eq:diphoton}).}
	\label{fig:diagrams}
\end{figure}

A pseudo-scalar particle with a mass $m_a \simeq 17$~MeV and a coupling to electrons of order $m_e/f_a \simeq 5\times 10^{-4}$ for $f_a \simeq 1$~GeV
contributes in a relevant way to the electron anomalous magnetic moment $a_e = (g-2)_e/2$. 
The measured deviation of the electron magnetic dipole moment~\cite{Hanneke:2008tm, Hanneke:2010au}~ from the SM prediction~\cite{Aoyama:2014sxa, Mohr:2015ccw} extracted from the fine structure constant measured in recent cesium 
recoil experiments~\cite{Parker:2018vye} is
\begin{equation}
\label{eq:oldalpha}
\Delta a_{e}\equiv a_{e}^{\rm exp} - a_{e}^{\rm SM}=(-88 \pm 36)\times 10^{-14}.
\end{equation}

However, more recently the deviation inferred from a new, independent measurement of the fine-structure constant has been reported as~\cite{Morel:2020dww} 
\begin{equation}
\label{eq:newalpha}
\Delta a^{\prime}_{e}=(48 \pm 30)\times 10^{-14}.
\end{equation}

In the present case, the largest contribution of the model to $\Delta a_{e}$ comes from the tree-level coupling of the axion to electrons, induced by the one-loop diagram in Fig.~\ref{fig:diagrams}
\begin{equation}
\Delta a_{e}^{\text{1-loop}}=-\frac{1}{8\pi^2}(g_{a}^{e})^2\int_{0}^{1}dx\frac{(1-x)^{3}}{(1-x)^{2} + x(m_{a}/m_{e})^{2}}.
\end{equation}

At the 2-loop level, however, the magnetic moment receives an additional contribution from a Barr-Zee type diagram proportional to the product $g_{a}^{e}g_{a}^{\gamma\gamma}$, where the effective coupling to photons is defined by Eq.~(\ref{eq:diphoton}). This contribution is given by

\begin{equation}
\Delta a_{e}^{\text{2-loop}}=g_{a}^{e}g_{a}^{\gamma\gamma}\frac{m_{e}}{\pi^{2}}f_{PS}\left[\frac{4\pi f_{\pi}}{m_{a}}\right] ,
\end{equation}
where

\begin{equation}
f_{PS}[z]=\int_{0}^{1}dx\frac{z^{2}/2}{x(1-x)-z^2}\log\left(\frac{x(1-x)}{z^2}\right),
\end{equation}
and we have taken an effective cutoff at the scale $4\pi f_{\pi}$ as in~\cite{Alves:2017avw}. Note that while we have chosen to work in the linear basis of the axion couplings to fermions, a similar calculation for $(g-2)_{\mu}$ has been presented in \cite{Bauer:2017ris}. There, one trades the linear coupling of the axion to fermions with a derivative coupling and a di-photon coupling. We have checked numerically that the corresponding result for $(g-2)_{e}$ in the non-linear basis agrees with our results in the linear basis. In fact, using the appendices in \cite{Bauer:2017ris}, we have verified that the chiral lagrangian obtained from the linear couplings matches that obtained in the non-linear basis due to the specific choice of interactions in our model.

\begin{figure}[htb]
	\centering
	\includegraphics[scale=0.85]{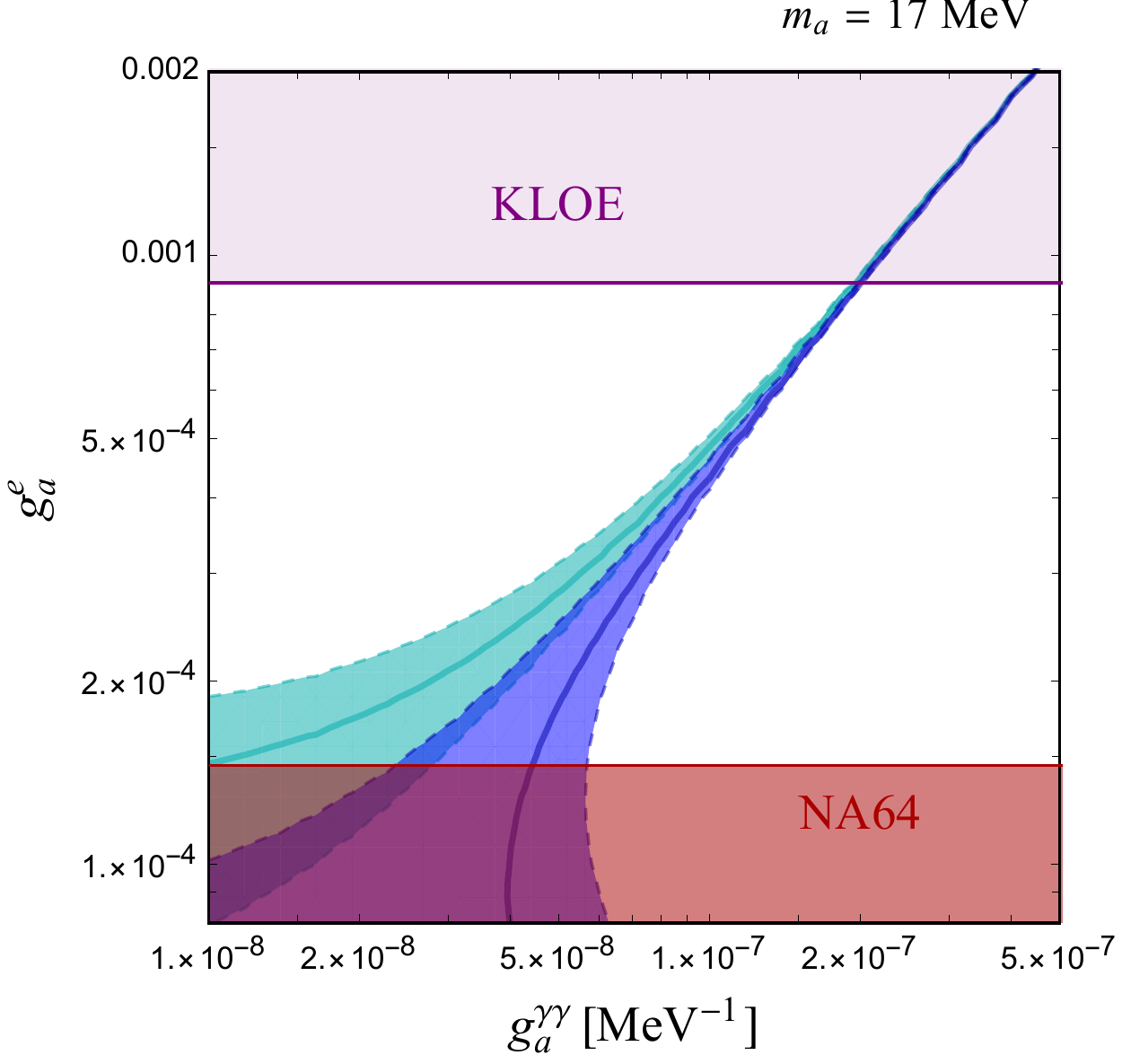}
	\caption{Parameter space for the 17 MeV axion which can accommodate $(g-2)_e$.  The purple and red shaded regions are excluded by KLOE and NA64 respectively~\cite{Liu:2018xkx,Banerjee:2019hmi}. The cyan region is consistent at the 95\% C.L. with the deviation $\Delta a_{e}$ obtained from Eq~(\ref{eq:oldalpha}). The blue region shows the 95\% C.L. region consistent with the value inferred from the recent measurement of the fine structure constant, reported in~\cite{Morel:2020dww}. }
	\label{fig:gm2_e}
\end{figure} 

It is clear that when $g_{a}^{e}$ and $g_{a}^{\gamma\gamma}$ have the same sign, the 2-loop contribution can partially cancel the 1-loop result. Thus, both couplings will have a relevant impact to the prediction of $ \Delta a_{e}$. In Fig.~\ref{fig:gm2_e}, we show the total contribution to $\Delta a_{e}$ for $m_{a}=17$ MeV and $f_{a}=1$~GeV with respect to the electron and photon effective couplings. The cyan region shows the range of couplings predicting $\Delta a_{e}$ at the 95\% CL according to the values reported in Eq.~(\ref{eq:oldalpha}), whereas the blue shaded region shows the region of parameters which can explain the new measurement of the deviation of the electron anomalous magnetic moment measurement, Eq.~(\ref{eq:newalpha}). 

Relevant constraints on the values of the electron coupling are extracted from the decays of the axion to electron-positron pairs. As these constraints are placed on the couplings to electrons, the constraints will remain the same regardless of which measurement of $\Delta a_{e}$ is assumed. Thus, in Fig.~\ref{fig:gm2_e} we show an upper limit on $g_{e}$ from the KLOE experiment in the purple shaded region~\cite{Liu:2018xkx}. A lower limit can be derived from various beam dump experiments. We show the strongest available bound, from NA64, in the red shaded region~\cite{Banerjee:2019hmi}. It is important to note that both bounds have been derived assuming ${\rm BR}(a\rightarrow e^{+}e^{-})=1$. We see that without the 2-loop contribution (or $g_{a}^{\gamma\gamma}\rightarrow 0$) the values of $\Delta a_e$ would be negative and therefore the new determination of the fine structure constant  demands sizable two loop corrections for the axion contributions to be consistent with $\Delta a_{e}$. Upper bounds on the couplings of the axion to vector bosons from LEP and LHC have also been explored in various realizations of axions and ALPS~\cite{Jaeckel:2015jla, Liu:2017zdh, Lee:2018lcj,Aaboud:2019ynx, Florez:2021zoo}. However, those bounds are based on the assumption of an axion which mostly decay into photons, contrary to our model in which the dominant axion decay is into electron pairs. Although searches for an axion decaying to electron pairs are difficult at LEP due to large backgrounds, this may be feasible at future runs of the LHC~\cite{Bauer:2018uxu}.

Let us stress that the values of the axion coupling to electron leads to a constraint on the ratio of  the PQ charges of the first generation quark and leptons.
For instance, since in this scenario, $Q_{u} \sim 2 Q_{d}$, from the expression of the axion mass one obtains that
\begin{equation}
\frac{f_a}{Q_{d}} \sim 1~{\rm GeV}.
\end{equation}
Therefore,  the coupling of the axion to the electrons, Eq.~(\ref{eq:fermioncoup}),  is given by
\begin{equation}
g_a^e = \frac{Q_{e}}{Q_{d}}  \frac{m_e}{{\rm GeV}} .
\end{equation}
Taking into account the range of values consistent with $(g-2)_e$,  which from Fig.~\ref{fig:gm2_e} are between $1.4 \times 10^{-4}$ and $1 \times 10^{-3}$, one obtains
\begin{equation}
\frac{Q_{e}}{Q_{d}} =  0.28 - 2.
\end{equation}

In particular, if the above ratio of PQ charges takes the values 1/3, 1/2 or 1, the axion is  free of experimental constraints from the KLOE and NA64 experiments. 
This range of values of the ratio of PQ charges will play an important role in the UV model described in section V.

\section{Atomki anomaly}
\label{sec:Atomki}

Recently, it has been shown that the variant of the QCD axion analyzed in this work, with couplings only to the first generation of quarks and leptons, can lead to an explanation of  the anomalies reported by the Atomki experiment~\cite{Alves:2020xhf}. The relevant axion-nuclear transition rates were also considered much earlier in~\cite{Treiman:1978ge,Donnelly:1978ty,Barroso:1981bp,Krauss:1986bq} with further studies aimed at generic pseudo-scalar couplings to nucleons more recently in~\cite{Cheng:2012qr,Ellwanger:2016wfe}. In terms of the mixing angles of the axion to mesons, the predicted axion emission rate for $^8\text{Be}^*(18.15)$ nuclear transitions can be parameterized as 

\begin{equation}
-(\theta_{a\eta_{ud}}(\Delta u + \Delta d) + \sqrt{2}\theta_{a\eta_{s}}\Delta s)\Big|_{^8\text{Be}^*(18.15)}\simeq (1.1 - 6.3)\times 10^{-4},
\label{eq:AtomkiBe}
\end{equation}
where $\Delta q$ parametrizes the proton spin contribution from a quark, $q$, while for $^4\text{He}^{*}(21.01)$ they obtain

\begin{equation}
-(\theta_{a\eta_{ud}}(\Delta u + \Delta d) + \sqrt{2}\theta_{a\eta_{s}}\Delta s)\Big|_{^4\text{He}^*(21.01)}\simeq (0.58 - 5.3)\times 10^{-4},
\label{eq:AtomkiHe}
\end{equation}
where the R.H.S of each equation gives the range of possible values explaining the Atomki anomalies.  For the aim of the computations, we have chosen $\Delta u + \Delta d = 0.52$ and $\Delta s = -0.022$, which
are consistent with the  lattice computations of these quantities~\cite{QCDSF:2011aa, Abdel-Rehim:2013wlz}. 

We note that since the axion-pion mixing must be small to remain acceptable from pion decay observations, the mixing angles $\theta_{a\eta_{ud}}$ and $\theta_{a\eta_{s}},$ which are determined by the Atomki anomaly, effectively fix the coupling of the axion to photons. We see in Fig.~\ref{fig:gm2_e} that for a given value of $g_{a}^{\gamma\gamma}$, the combined one- and two-loop contributions to $(g-2)_{e}$ sets a value for $g_{e}$ which leads to the correct prediction for $\Delta a_{e}$.  Thus, combining the regions of mixing angles which favor the Atomki anomalies with the requirement to satisfy $(g-2)_{e}$ parametrically connects the PQ charges of the $u$ and $d$ quarks to that of the electron, giving complementary information on the viable region of the scenario. 

If SM effects are confirmed to be dominant feature of the Atomki anomaly, as in~\cite{Aleksejevs:2021zjw}, this could imply further constraints on the mixing angles $\theta_{a\eta_{ud}}$ and $\theta_{a\eta_{s}}$ or more specific relations between them. For instance, in order to suppress the resonant signal, one 
possibility would be that  those mixing angles are much smaller than their natural values, leading to a small photon axion coupling. Beyond the fact that such small mixing values are difficult to realize, a small photon axion coupling would be inconsistent with the values of $\Delta a_e$ associated with the new fine structure constant determination, Eq.~(\ref{eq:newalpha}), but would remain consistent with the old one, Eq.~(\ref{eq:oldalpha}).  Alternatively, there could be an approximate cancellation between the   $\theta_{a\eta_{ud}}$ and $\theta_{a\eta_{s}}$  contributions to Eqs.~(\ref{eq:AtomkiBe}) and (\ref{eq:AtomkiHe}), what
would imply a mixing $\theta_{a\eta_{s}}$ about an order of magnitude larger than $\theta_{a\eta_{ud}}$ .

\begin{figure}[t]
\centering
\includegraphics[scale=0.6]{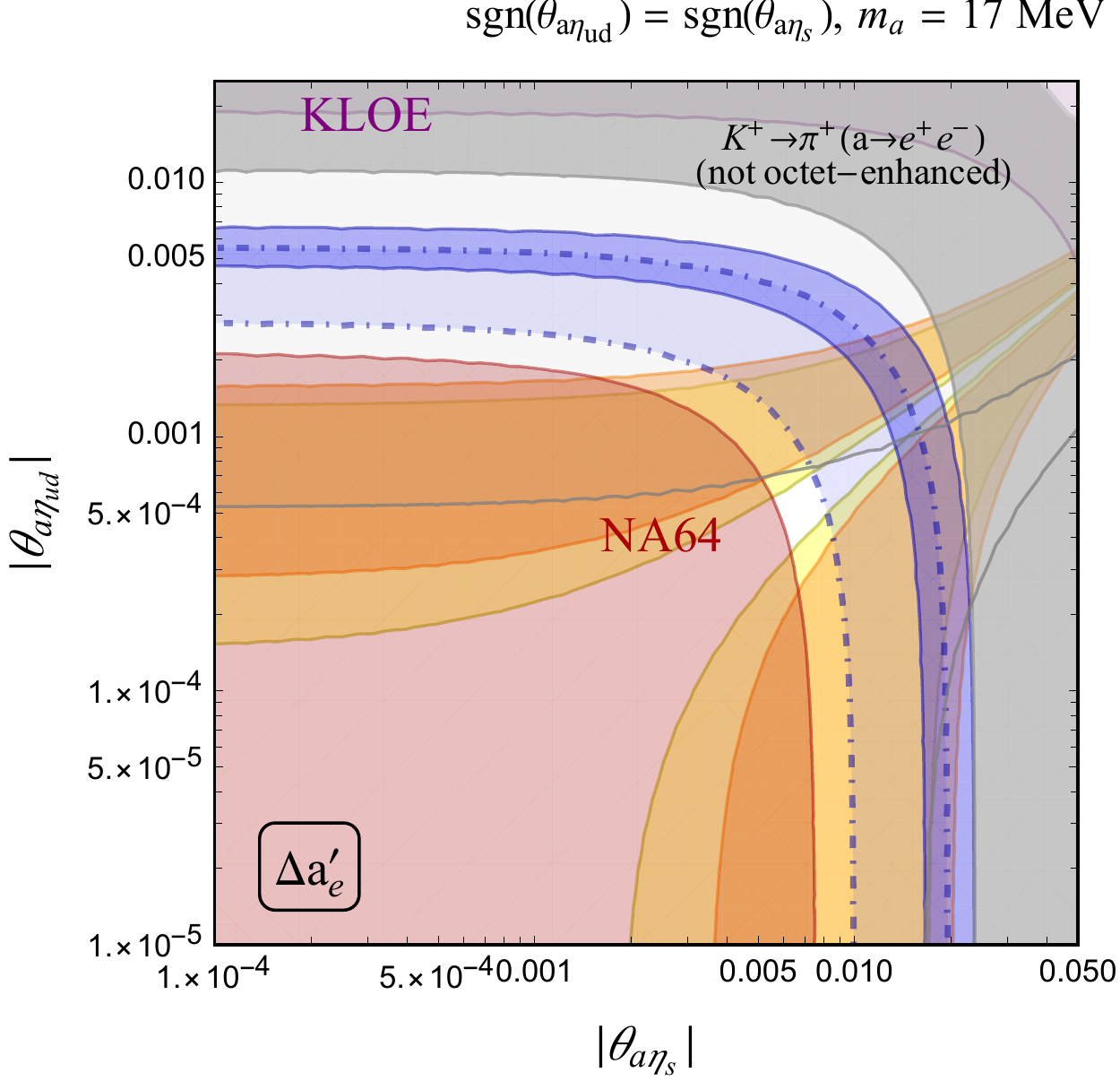}
\includegraphics[scale=0.6]{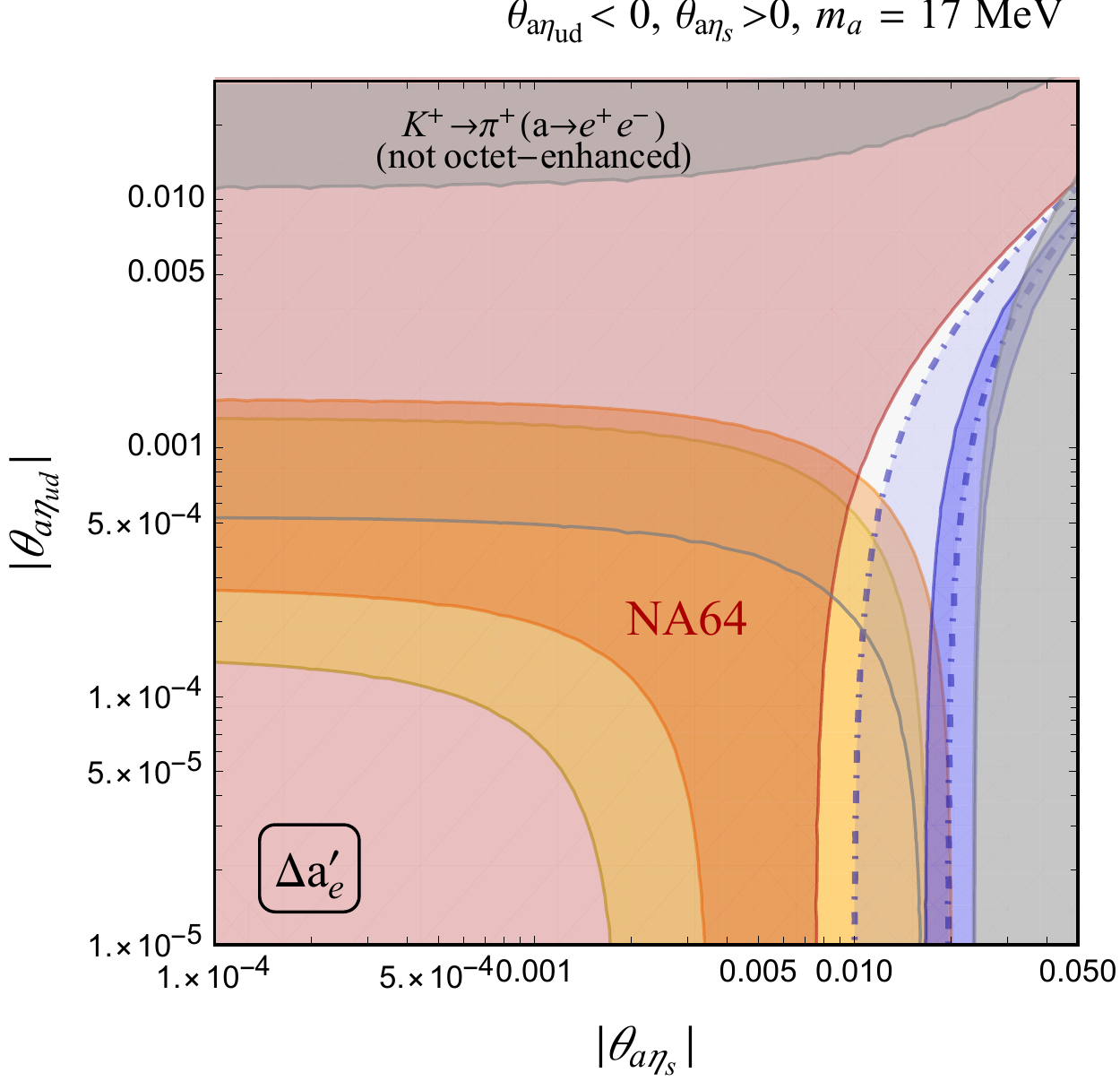}
\caption{Mutual parameter space for the $(g-2)_{e}$ and Atomki anomalies. {\bf{Left:}} The light and dark yellow bands indicate the parameter space for the $17$ MeV axion which can accommodate the Atomki anomalies for ${}^8$Be and ${}^4$He, respectively, when $\text{sgn}(\theta_{a\eta_{ud}}) = \text{sgn}(\theta_{a\eta_{s}})$. We show exclusion regions from NA64 and KLOE in red and purple respectively, as in Fig. \ref{fig:gm2_e}, assuming the correlation of electron couplings consistent with $\Delta a^{\prime}_{e}$ at the 95\% C.L. Light gray regions are excluded from kaon decays in the non octet-enhanced regime. Gray lines indicate the region that would be excluded in the octet-enhanced regime.  {\bf{Right:}} Same parameter space when $\theta_{a\eta_{ud}} <0$ and  $\theta_{a\eta_{s}}>0$. In both panels, the light (dark) blue shaded contours show regions where $g_{a}^{\gamma\gamma}$ is consistent with $\Delta a^{\prime}_{e}$ at the 95\% C.L. for $Q_{e}/Q_d=1/3~(1/2)$.}
\label{fig:atomki_1}
\end{figure} 

In Fig.~\ref{fig:atomki_1} we show the regions of parameters given in \cite{Alves:2020xhf} favored by the nuclear transition anomalies. We include the correlation between the predicted value of $g_{a}^{\gamma\gamma}$ in this plane with the values of $g_{e}$ which are consistent with $\Delta a^{\prime}_{e}$, Eq.~(\ref{eq:newalpha}),  at the 95\% CL. Thus, the red and purple shaded regions are excluded by the NA64 and KLOE experiments, respectively, as shown in Fig. \ref{fig:gm2_e}. The light gray region is excluded by bounds on the $K^{+}\rightarrow \pi^{+}(a\rightarrow e^{+}e^{-})$ branching ratio in the non octet-enhanced  regime, see~\cite{Alves:2020xhf}. The regions that would be excluded in the octet enhanced regime, which are therefore not firm bounds,  are indicated by gray lines. In the left panel, we show the predictions and bounds for the case when $\text{sgn}(\theta_{a\eta_{ud}}) = \text{sgn}(\theta_{a\eta_{s}})$, while in the right panel we show the corresponding results when $\theta_{a\eta_{ud}} <0$ and  $\theta_{a\eta_{s}}>0$.\footnote{When $\theta_{a\eta_{ud}} >0$ and  $\theta_{a\eta_{s}}<0$ we find that there is no mutual parameter space for the $(g-2)_{e}$ and Atomki anomalies.} 

\begin{figure}[t]
\centering
\includegraphics[scale=0.6]{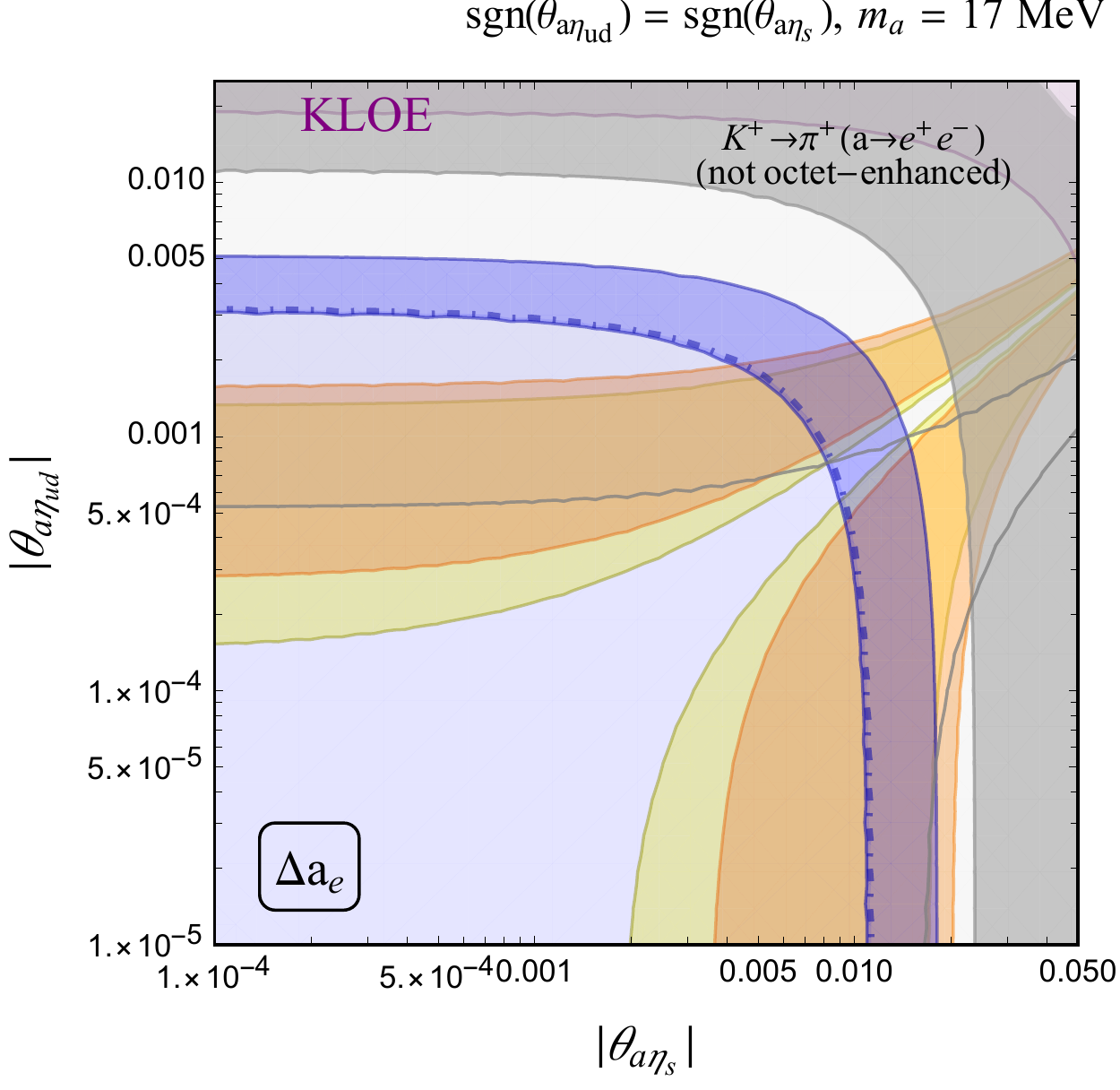}
\includegraphics[scale=0.6]{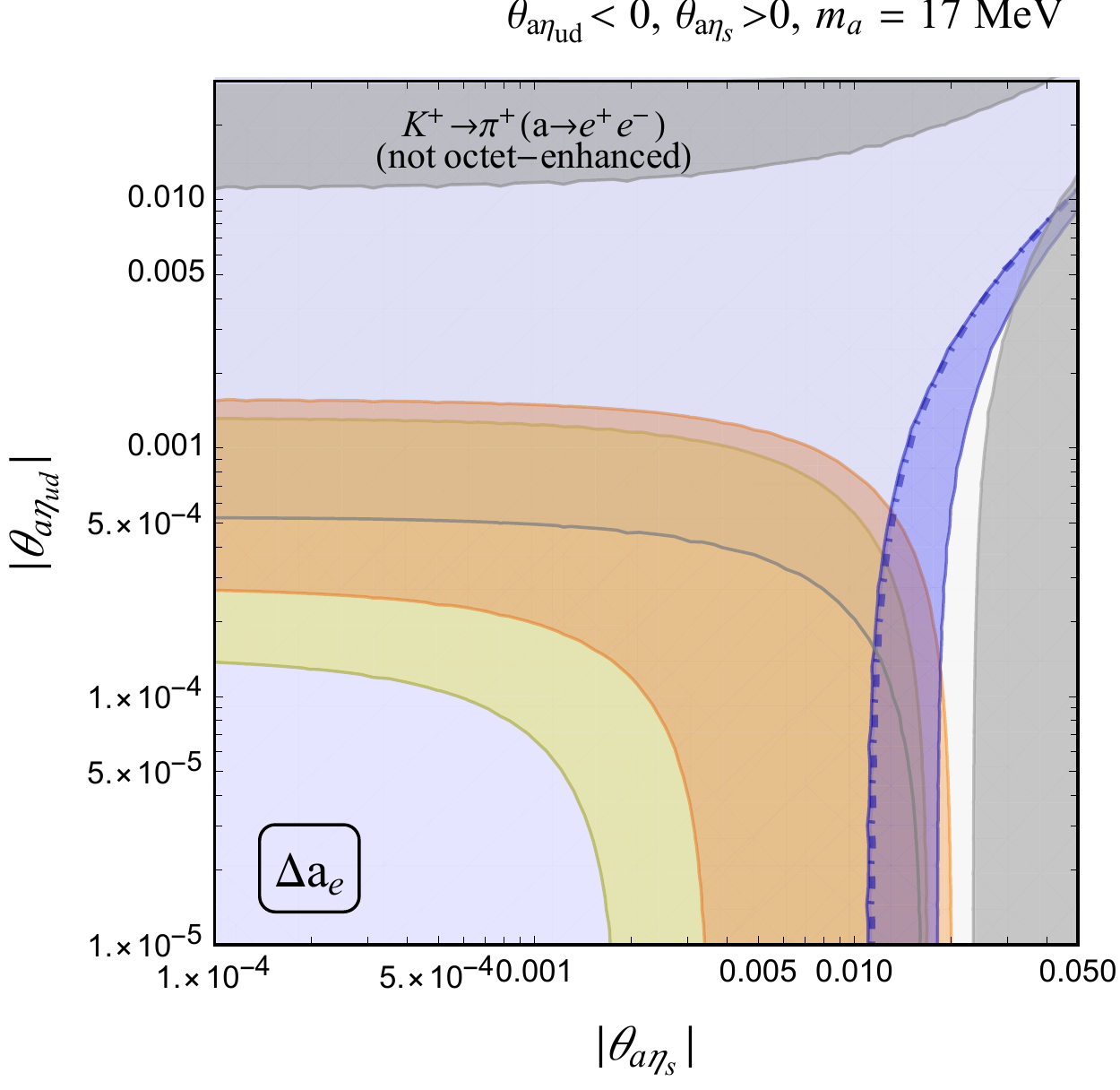}
\caption{Same parameter space as in Fig.~\ref{fig:atomki_1} assuming that the correlation of $g_{e}$ and $g_{a\gamma\gamma}$ is consistent with the Berkeley measuremant, $\Delta a_{e}$, at the $95\%$ CL. The bound from NA64 inferred from the central value of $\Delta a_{e}$ is shown in the red dashed line.}
\label{fig:atomki_2}
\end{figure}

It would be interesting to compare the allowed values of the axion-$\eta$ mixing parameters with the ones predicted in chiral perturbation theory. However, as emphasized in Ref.~\cite{Alves:2017avw}, this is not a
simple task, since these mixing parameters receive relevant contributions at next-to-leading order. The only firm prediction is that these mixing parameters are naturally between $10^{-3}$ and a few $10^{-2}$, and
hence, they are consistent with most of the allowed parameter space shown in Fig.~\ref{fig:atomki_1}. Between the yellow regions the contributions from $\theta_{a\eta_{ud}}$ and  $\theta_{a\eta_{s}}$ cancel in Eqs.~(\ref{eq:AtomkiBe}) and (\ref{eq:AtomkiHe}) leading to a suppression of the axion signal in the $^8\text{Be}^*(18.15)$ and $^4\text{He}^{*}(21.01)$ transitions.

The light and dark blue bands in Fig.~\ref{fig:atomki_1} also show the regions of parameter space consistent with a ratio of PQ charges $Q_{e}/Q_{d} = 1/3$ and 1/2, respectively. We see that for opposite signs of the mixing angles, as represented in the right panel of Fig.~\ref{fig:atomki_1}, a ratio $Q_e/Q_d = 1/2$ leads to a mixing angle $\theta_{a \eta_s}$ that is much larger than its natural values. 

The corresponding parameter space consistent with the determination of $\Delta a_{e}$ from Eq.~\eqref{eq:oldalpha} is, however, quite different compared to that of Eq.~\eqref{eq:newalpha}. In Fig.~\ref{fig:gm2_e} we see that, except for values of $\Delta a_{e}$ close to or larger than the central value of the measurement, there is no lower bound on $g_{a\gamma\gamma}$ inferred from NA64. Hence, this constraint on the meson-mixing angles vanishes for couplings consistent with the region $[\Delta a_{e}-2\sigma,\Delta a_{e}]$. In Fig.~\ref{fig:atomki_2}, we show the corresponding parameter space for the $(g-2)_e$ and Atomki anomalies assuming the correlation of electron and photon couplings consistent with $\Delta a_{e}$, from Eq.~\ref{eq:oldalpha}, at the $95\%$ CL. All color shading follows that of Fig.~\ref{fig:atomki_1}. In this case, natural regions of $\theta_{a\eta_{ud}}$ and $\theta_{a\eta_{s}}$ accounting for both anomalies are completely open. In fact, scenarios motivated by PQ charges $Q_e/Q_d = 1/3$ cover almost this whole region. However, this is not the case for scenarios motivated by $Q_e/Q_d = 1/2$ which does not differ significantly considering either determination of the electron magnetic moment.

\section{UV model}
\label{sec:model}

\subsection{The generation of CKM matrix and the couplings to first generation fermions}

In this section, we present a possible UV completion of the effective model presented in Eq.~(\ref{eq:lag}), where the axion couples exclusively to the first generation fermions in the SM. Generating the effective interactions to the up- and down-quarks is non-trivial as the PQ breaking mechanism must be carefully intertwined with electroweak symmetry breaking in a way that reproduces the correct flavor structure in the CKM matrix. To this end, we consider an extension of the SM by three additional Higgs doublets and three singlets Higgs fields. The PQ symmetry is realized by assigning charges to the additional Higgs bosons as well as the right-handed SM fermions $u_{R}$, $d_{R}$, and $e_{R}$. The particle content and their PQ charges of the model is summarized in Table~\ref{tab:charges}.

\begin{table}[h]
	\begin{tabular}{|c|c|c|c|c|c|c|c|}
		\hline 
		Particles  & $H$ & $H_u$ & $H_{d,e}$& $u_R $ & $d_R$ & $e_R$ & $\phi_f$ \\  \hline
		\hline
		$SU(2)_L$    & 2 & 2 & 2 & 1  & 1  &1 & 1  \\
		\hline
		$U(1)_Y$      & $\frac{1}{2}$   & $-\frac{1}{2}$   &$\frac{1}{2}$& $\frac{2}{3}$    & $-\frac{1}{3}$ & $-1$ & $0$ \\
		\hline
		$U(1)_{\rm PQ}$  & 0  & $-Q_u$ & $-Q_{d,e}$ &  $Q_u$   & $Q_d$ & $Q_e$ & $-Q_f$ \\
		\hline
	\end{tabular}
	\caption{ $ SU(2)_L \times U(1)_Y \times U(1)_{\rm PQ}$ charges of the SM Higgs, additional Higgs doublets $H_{f}$, right-handed fermions, and Higgs singlets $\phi_{f}$, where $U(1)_{\rm PQ}$ is a global Peccei-Quinn-like symmetry and $f = u,d,e $.
		The PQ charges are $Q_u = 2$, $Q_d = 1$ and $Q_e = 1/n$, with $n=2$ or $3$. All other particles are considered to be PQ singlets.}
	\label{tab:charges}
\end{table}

The relevant Yukawa interactions allowed by the PQ symmetry of the Higgs doublets, $H_{f}$, and those of the SM Higgs, $H$, are given by 
\begin{align}
	\mathcal{L}_{\rm PQ}^{\rm Yuk}  \supset & - \sum_{i=1,2,3}\left(\bar{Q}^i Y^{i1}_u H_u u_R^1 + \bar{Q}^i Y^{i1}_d H_d d_R^1  + \bar{L}^i Y^{i1}_e H_e e_R^1\right) +h.c. \\
	\mathcal{L}_{\rm SM}^{\rm Yuk}  \supset & - \sum_{i=1,2,3}\sum_{j=2,3}\left(\bar{Q}^i Y^{ij}_u \tilde{H} u_R^j + \bar{Q}^i Y^{ij}_d H d_R^j 
	+ \bar{L}^i Y^{ij}_e H e_R^j \right) +h.c. ,
\end{align}
where $\mathcal{L}_{\rm PQ}^{\rm Yuk}$ gives the couplings of $H_{u,d,e}$ to the first generation fermions, while $\mathcal{L}_{\rm SM}^{\rm Yuk} $ gives the couplings of the second and third generation fermions to the SM Higgs. Below the scales of PQ and electroweak symmetry breaking the up-quark mass matrix is then given by
\begin{align}
	M_u \equiv \frac{1}{\sqrt{2}}
	\left(\begin{array}{ccc}
		Y_u^{11} v_{u} &  Y_u^{12} v &  Y_u^{13} v \\
		Y_u^{21} v_{u} &  Y_u^{22} v &  Y_u^{23} v \\
		Y_u^{31} v_{u} &  Y_u^{32} v &  Y_u^{33} v \\
	\end{array}\right)
	=  V_{u_L}
	\left(\begin{array}{ccc}
		m_u  &  0 &  0 \\
		0 &  m_c &  0 \\
		0 &  0 &  m_t \\
	\end{array}\right) V_{u_R}^\dagger
\label{eq:upmasses}
\end{align}
where $v$ and $v_u$ are the vacuum expectation values of $H$ and $H_u$ respectively. The $V_{uL}$ and $V_{uR}$ matrices given the rotation between the flavor- and mass-eigenstate basis (denoted with superscript $m$),
\begin{align}
	\vec{u}_L = V_{uL} \vec{u}_L^{\; m}, \quad
		\vec{u}_R = V_{uR} \vec{u}_R^{\; m}, 
\end{align}
where $\vec{u}_{L/R}=(u_{L/R}, c_{L/R}, t_{L/R})^{T}$. For simplicity, we work in a basis where $V_{u_R}$ is the identity matrix. The diagonalization of the down-type quarks follows similarly. 

After diagonalization of both up- and down-type quark sectors, the CKM matrix is given by
\begin{align}
V_{\rm CKM} = V_{u_L}^\dagger V_{d_L} .
\label{eq:CKM}
\end{align} 
Since $H_u$ is charged under the PQ symmetry while $H$ is not, we expect the axion field
to be contained in $H_u$ but not $H$.
Therefore, to connect to the effective model, this requires that $H_u$ couples only to the first generation up-quark in the mass eigenstate basis, 
\begin{align}
\bar{Q}^i Y^{i1}_u H_u u_R^1 =\overline{Q^{m}}^i \left(V_{u_L}^\dagger \right)^{ij} Y^{j1}_u H_u u_R^{1,m}
= \sqrt{2} \frac{m_u}{v_{u}}  \overline{Q^{m}}^1  H_u u_R^{1,m}  ,
\end{align} 
which is equivalent to 
\begin{align}
	\left(V_{u_L}^\dagger \right)^{ij} Y^{j1}_u =\sqrt{2} \frac{m_u}{v_{u}}   \left(1, 0, 0 \right)^T.
	\label{eq:req}
\end{align}
Taking $Y^{j1}=\left(Y^{11}_u, Y^{21}_u, Y^{31}_u \right)^T$, the unitary matrix $V_{u_L}$ can be decomposed in terms of the normalized vector $(Y^{j1})_{\parallel}$,
and two perpendicular normalized vectors $(Y^{j1})_{\bot 1}$ and $(Y^{j1})_{\bot 2}$
\begin{align}
	V_{u_L}^\dagger = \left(\begin{array}{ccc}
	1  &  0 &  0 \\
	0 &  c_x &  s_x \\
	0 &  -s_x &  c_x \\
	\end{array}\right)
	\left(\begin{array}{c}
	(Y^{j1}_u)_{\parallel}^T \\ (Y^{j1}_u)_{\bot 1}^T   \\  (Y^{j1}_u)_{\bot 2}^T 
	\end{array}\right),
\label{eq:Vu}
\end{align}
where $T$ represents the transposed column vector, and $x$ is an auxiliary rotation angle which does not affect the equality Eq.~(\ref{eq:req}). Similarly for $V_{d_{L}}$ we obtain
\begin{align}
	V_{d_L}^\dagger = \left(\begin{array}{ccc}
	1  &  0 &  0 \\
	0 &  c_y &  s_y \\
	0 &  -s_y &  c_y \\
\end{array}\right)
\left(\begin{array}{c}
	(Y^{j1}_d)_{\parallel}^T \\ (Y^{j1}_d)_{\bot 1}^T   \\  (Y^{j1}_d)_{\bot 2}^T .
\end{array}\right). 
\label{eq:Vd}
\end{align}

We choose the auxiliary angles explicitly as
\begin{align}
	& x = \theta_{23}, \quad y = 0, 
\end{align}
and for the Yukawas we take
\begin{align}
	& \left(\begin{array}{c}
		(Y^{j1}_u)_{\parallel}^T \\ (Y^{j1}_u)_{\bot 1}^T   \\  (Y^{j1}_u)_{\bot 2}^T 
	\end{array}\right) = 
 \left(\begin{array}{ccc}
	c_{13}  &  0 &  s_{13} \\
	0 &  1 &  0 \\
	-s_{13} &  0 &  c_{13} \\
\end{array}\right) , \quad
	\left(\begin{array}{c}
	(Y^{j1}_d)_{\parallel}^T \\ (Y^{j1}_d)_{\bot 1}^T   \\  (Y^{j1}_d)_{\bot 2}^T 
\end{array}\right) = 
\left(\begin{array}{ccc}
	c_{12}  &  - s_{12} &  0 \\
	s_{12} &  c_{12} &  0 \\
	0 &  0 &  1 \\
\end{array}\right),
\label{eq:explicit}
\end{align}
where $\theta_{12}$, $\theta_{13}$, $\theta_{23}$ angles are  the CKM mixing angles in the SM. Then, the explicit forms of the vectors  $Y_u^{j1}$ and $Y_d^{j1}$ are given by,
\begin{align}
	Y_u^{j1} = \sqrt{2} \frac{m_u}{v_{u}} 
	\left(\begin{array}{c}
		c_{13}  \\ 0 \\  s_{13}
	\end{array}\right), \quad
	Y_d^{j1} = \sqrt{2} \frac{m_d}{v_{d}} 
	\left(\begin{array}{c}
		c_{12}  \\ -s_{12} \\  0
	\end{array}\right).
\end{align}
The rest of the Yukawa matrix $Y_u$ and $Y_d$ can be reconstructed according to Eq.~(\ref{eq:upmasses}). This completes the realization of the CKM matrix in the UV model. We omit details of the procedure of the mixing in the lepton sector. However, this may be accomplished in an analogous way.

Besides the Yukawa interactions, the scalar potential needs to be studied to single out the physical axion field in the model as well as its mixing with other pseudo-scalars.
The off-diagonal scalar potential, $V_{\rm PQ}$, and diagonal scalar potential, $V_{\rm dia}$, allowed by symmetries are 
\begin{align}
	V_{\rm PQ}  & =
	\left(A_u \phi_u^* H \cdot H_u  + A_d \phi_d^* H^\dagger H_d  + A_e \phi_e^* H^\dagger H_e 
	+ A_\phi \phi_u^* \phi_d^2 + B_\phi \phi_d^* \phi_e^n \right)  + h.c. ,
	\label{eq:VPQ}
	\\
	V_{\rm dia}  &= \sum_{\Phi}
	- \mu_\Phi^2 \Phi^\dagger  \Phi + \lambda_\Phi \left(\Phi^\dagger  \Phi \right)^2,
	\label{eq:Vdia}
\end{align}
where $\Phi= H, H_u, H_d, H_e, \phi_u, \phi_d, \phi_e$. 
We assume that all the coefficients are real. In Eq.~(\ref{eq:VPQ}), the first three terms generate the couplings of
axion to first generation fermions through mixing with $H_{u,d,e}$. 
The last two terms reflect the PQ charge assignments of the scalars $\phi_u$, $\phi_d$ and $\phi_e$. $V_{\rm dia}$ provides all the diagonal terms which do not provide mixing, but provide masses to CP-even scalars. 
We omit other operators found by other combinations of the scalar fields allowed by the imposed symmetries. The above potential will be sufficient for a complete description of the physical axion and the interactions relevant for our discussion.

We require that the potential has a minimum, which determines the values of the $\mu_{\Phi}$'s. After applying this condition,
we solve the mixing between the seven pseudo-scalars, $\vec{\Phi}^I \equiv (h_I, -h_u^I, h_d^I, h_e^I, \phi_u^I, \phi_d^I,\phi_e^I)^{T}$, of which five will be massive, where we have chosen $-h_u^I$ since it is associated with the imaginary part of the neutral component of $\tilde{H}_u$, carrying the same hypercharge as $H_d$ and $H_e$. The two massless pseudoscalars correspond to
the Goldstone bosons after breaking of $SU(2)_L \times U(1)_Y$ ($\vec{G}_{SM}$ eigenvector) and the global $U(1)_{PQ}$ ($\vec{G}_{\rm PQ}$ eigenvector) symmetries
\begin{align}
	\vec{G}_{\rm SM} &	= \frac{1}{\sqrt{v^2+v_u^2+v_d^2 + v_e^2} } \left(\begin{array}{ccccccc}
		v,  &  v_u, &  v_d, & v_e & 0, & 0, & 0 
	\end{array}\right), \\
	\vec{G}_{\rm PQ}	& \approx \frac{1}{\sqrt{ \sum_{f} Q_f^2( v_f^2 + v_{\phi_f}^2)  } } \times \nonumber\\
	& \left(\begin{array}{ccccccc}
		-\frac{ \sum_f  (-1)^{f} Q_f  v_f^2  }{v},  &  -Q_u v_u, & Q_d v_d, &Q_e v_e,
		& Q_u v_{\phi_u}, & Q_d v_{\phi_d}, & Q_e v_{\phi_e}
	\end{array}\right), 
\end{align}
where $f = u,d,e$, $Q_f$ is the associated PQ charge, and we define $(-1)^{f}\equiv-1$ for $f=u$ and $+1$ for $f=d,e$. 
We have expressed $\vec{G}_{\rm PQ}$ at leading order in $v \gg v_{f}, v_{\phi_f}$.

Note that $\vec{G}_{\rm PQ}$ gives the mixing between the physical axion states among the seven pseudo-scalars, e.g.
$\vec{\Phi}^I \supset \vec{G}_{\rm PQ} a$. 
For example, if we further assume $v_{\phi_f} \gg v_f$, it is clear that the axion is dominantly composed of $\phi_f^I$.
In the doublet Higgs $H_f$, the axion is contained with the suppressed factor $v_f/v_{\phi_f}$,
while for the SM Higgs $H$, it is contained with a double suppressed factor $v_f^2/(v v_{\phi_f})$.

After rotating to the physical basis, the couplings of axion to SM fermions are given by
\begin{align}
	\sum_{f={u,d,e}} \frac{m_f}{f_a} Q_f i a \bar{f}\gamma_5 f 
	- \frac{ \sum_f  (-1)^{f} Q_f  v_f^2 }{v^2} \sum_{F = {\rm 2nd, 3rd}} \frac{m_F}{f_a} i a \bar{F} \gamma_5 F,
	\label{eq:axionffcoupling}
\end{align}
where $f_a \equiv \sqrt{  \sum_{f} Q_f^2( v_f^2 + v_{\phi_f}^2)  }$ and the second term gives the axion couplings to the 2nd and 3rd generation fermions, generated by mixing with the SM Higgs. Note that this generates an effective PQ charge to heavy fermions, namely
\begin{align}
	Q_{F,~{\rm eff}}^{\rm PQ} \equiv - \frac{ \sum_f  (-1)^{f} Q_f  v_f^2}{v^2}.
\end{align}

The couplings of the QCD axion to the 2nd and 3rd generation fermions are constrained from the exotic decays of  heavy mesons.
The relevant constraints are $J/\Psi \to \gamma a$ leading to $|Q_c^{\rm PQ}| \lesssim 0.25 (f_a/{\rm GeV})$
and  $\Upsilon \to \gamma a$ leading to $|Q_b^{\rm PQ}| \lesssim 0.8\times 10^{-2} (f_a/{\rm GeV})$ \cite{Alves:2017avw}.
If we take the hierarchy $v_f \approx 100~{\rm MeV} \ll v$ and $f_a\approx 1$ GeV, we have 
$Q_{F,~{\rm eff}}^{\rm PQ} \approx 5 \times 10^{-7}$ which is quite safe from the above constraints.

\subsection{Additional Massive Scalar and Pseudo-scalar States}

In addition to the two Goldstone bosons, there are five massive pseudo-scalars. 
For $n =3$, e.g. $Q_e^{\rm PQ} = 1/3$,
one can only decouple four heavy massive pseudo-scalars leaving one light pseudo-scalar.
More specifically, we assume the following parameters
\begin{align}
	A_\phi \gg A_{u,d,e} \equiv A_f \approx 20~{\rm GeV}, ~ v_{u,d,e} \equiv v_f \approx {\rm 20 ~MeV}, ~ 
	v_{\phi_{u,d,e}}\equiv v_{\phi_f} \approx {\rm 1 ~GeV},
\end{align}
where the last term determines the axion decay constant $f_a \approx \mathcal{O}(1) {\rm GeV}$. 
In this case, $B_\phi$ is a dimensionless parameter.  There are then  three massive pseudo-scalars, composed mostly of linear combinations of the $h_f^I$ states, with masses given by
\begin{align}
	A_f v v_{\phi_f}/v_f,
\end{align}
where $f=u,d,e$. Additionally, there is one mass of the form 
\begin{align}
 	A_\phi v_{\phi_f},
\end{align}
which dominantly comes from linear combination of $\phi_{u,d}^I$ states and may have a mass close to 100~GeV, and one pseudo-scalar of mass
\begin{align}
 \quad  B_\phi v_\phi^2.
 \end{align}

The light pseudo-scalar with mass $B_\phi v_\phi^2 \sim {\rm GeV}^2$, which we denote as $\phi_{\rm light}^I$, mainly comes from $\phi_e^I$. 
The lightness of this state can be understood in the limit $B_\phi =0$. In this case, there is a global $U(1)_e$ symmetry and associated charged scalars $\phi_e$ and $H_e$. As a result, there is one additional exact goldstone boson.
With $B_\phi \neq 0$, the $U(1)_e$ is explicitly broken in the scalar potential. Therefore, $\phi_{\rm light}^I$ becomes a pseudo-goldstone boson with mass proportional to $B_\phi v_{\phi_f}^2$ and not related to any other dimensionful trilinear couplings. 
The associated eigenvector can be easily represented with the simplficiation $A_\phi = A_u = 0$. In this case, we obtain
\begin{align}
	\vec{\omega} (\phi_{\rm light}^I) & \approx \left\{ 0,0,v_d (v_e^2 + v_{\phi_e}^2), -3 v_e (v_d^2 + v_{\phi_d}^2),
	0, v_{\phi_d} (v_e^2 + v_{\phi_e}^2),  -3 v_{\phi_e} (v_d^2 + v_{\phi_d}^2)\right\}, \nonumber \\
	& \approx \left\{ 0,0,v_f, -3 v_f,
	0, v_{\phi_f} ,  -3 v_{\phi_f} \right\},
\end{align}
where we have omitted the normalization factor. Thus, $\phi_{\rm light}^I$ is dominantly composed of $\phi_e^I$ since $v_f \ll v_{\phi_f}$. Moreover, it has a mixing of $v_f/v_{\phi_f}$ from $H_e^I$ and an even smaller mixing from $H_d^I$.
Therefore, this state mainly couples to the electron with a coupling of $\sim m_e/v_\phi \sim 5\times 10^{-4}$.
This coupling can be constrained using the Dark photon search at BaBar, $e^+ e^- \to \gamma A'$~\cite{Lees:2014xha}. Refs.~\cite{Knapen:2017xzo, Jana:2020pxx} have converted this constraint to scalars, which requires a coupling smaller than $\sim 5\times 10^{-4} $ for scalar masses $1$--$10$ GeV, and therefore this light pseudo-scalar coupling would be close to the current experimental bounds. The corresponding 1-loop contribution to $\Delta a_e$ is $-1.1\times 10^{-14}$ (taking $m_{\phi_{\rm light}^I}=1$ GeV and coupling $5\times 10^{-4}$) making its contribution to $\Delta a_e$ much smaller than the uncertainty of the
experiment. 

For $n=2$, $B_\phi$ is dimensionful. 
In this case, the mass of $\phi_{\rm light}^I$ is a linear combination of $A_\phi v_\phi$ and $B_\phi v_\phi$.
One can choose $A_\phi$ and  $B_\phi$ large enough such that $\phi_{\rm light}^I$ avoids existing constraints.
Moreover, one can also add the term $A_1 \phi_e H^\dagger_d H_e$, which can also further raise its mass.
As a result, we conclude that for $n =3$ the light pseudo-scalar is still viable but is subject to strong constraints, while for $n=2$, the five massive pseudo-scalars can be heavy and completely decoupled from the low energy phenomenology.

Having dealt with the pseudo-scalar sector, we briefly discuss the seven massive CP-even scalars. 
For $n=3$, there are seven massive CP-even scalars.
Under the assumption $A_f \gg v_{\phi_f} \gg v_f$, we find that there is one SM-like Higgs with mass squared $2 \lambda_H v^2$. In this limit, there are three CP-even masses squared  given by
\begin{align}
	 ~ A_f v \frac{v_{\phi_f}}{v_f},
\end{align}
where $f=u,d,e$, and are all on the order of $\sim (500~{\rm GeV})^2$. These states are mainly composed of the neutral CP-even components of doublet Higgs $H_f$.
There are two massive scalars  with masses squared $\sim A_{\phi} v_{\phi_f} \simeq (100~{\rm GeV})^2$, which are mostly composed of the CP-even components of singlet Higgs bosons $\phi_f$.\footnote{Since $v_{\phi_f}=1$ GeV, this requires values for $A_{\phi}$ which may induce a second minimum in the scalar potential away from the EW one, particularly in the direction where the doublet fields vanish. In this case, the EW vacuum must be in a metastable configuration just as in the SM. However, as mentioned previously we have omitted additional terms in Eq.(\ref{eq:VPQ}) which are irrelevant to the main issues in this work and a full evaluation of the potential and its vacuum structure is beyond the present discussion.}
Similar to the pseudo-scalar sector, there is still one light scalar with mass squared $\sim \mathcal{O}(v_{\phi_f})^2$, whose eigenvector is given by
\begin{align}
	\vec{\omega} (\phi_{\rm light}^R) & \approx \left\{ \frac{A_e v_e v_{\phi_e}}{m^2_{h, {\rm SM} } \sqrt{v_e^2+v_{\phi_e}^2} },0,0,
	\frac{v_e}{\sqrt{v_e^2+v_{\phi_e}^2} },
	0, 0, \frac{v_{\phi_e}}{\sqrt{v_e^2+v_{\phi_e}^2}}\right\}.
\end{align}
Just as $\phi^I_{\rm light}$, $\phi_{\rm light}^R$ couples dominantly to the electron with coupling $\sim m_e/v_{\phi_e}$ and suffers from similar constraints.
For $n=2$, the light scalar masses become $A_\phi v_\phi $ or $B_\phi v_\phi$. Thus, in this case the masses for CP-even scalars are also heavy and can be decoupled from low energy phenomenology.

Finally, we come to the phenomenology related to the SM Higgs decay. Working again in the limit $A_f \gg v_{\phi_f} \gg v_f$ the mass eigenstate of the SM Higgs is almost exactly given by the neutral CP-even component of $H$. The interactions which determine the decay of the SM Higgs to two axions are given by
\begin{align}
	\mathcal{L}_{\rm int}  & = h \left(\lambda v h_I^{2} - \frac{1}{\sqrt{2}} \sum_{f=u, d,e} (-1)^{f} A_f h_f^I \phi_f^I \right) \\
	& \supset - \frac{\sqrt{2}}{2} \frac{ \sum_{f=u,d,e} (-1)^{f}  Q_f^2 A_f v_f v_{\phi_f} }{ \sum_{f=u,d,e} Q_f^2 (v_f^2 + v_{\phi_f}^2 )  } h a^2 \approx - \frac{\sqrt{2}}{2} A_f \frac{v_{f}}{ v_{\phi_f} } h a^2,
\end{align}
where $f= u,~d,~e$, and $(-1)^{f}\equiv-1$ for $f=u$ and $+1$ for $f=d,e$, as defined above. We also assume $v_f \ll v_{\phi_f}$. The SM Higgs decay to $aa$ has a width of
\begin{align}
	\Gamma(h\to aa) = \frac{A_f^2 v_f^2}{16\pi m_h v_{\phi_f}^2}\sqrt{1 - 4\frac{m_a^2}{m_h^2}}\approx 0.025 ~{\rm MeV},
\end{align} 
which is far smaller than SM Higgs total width of $4$ MeV (leading to  ~$0.6\%$ exotic Higgs branching ratio). Assuming the boost factor for the axion is $m_h/(2 m_a)$ and width of $a \to e^+ e^-$ as
\begin{align}
	\Gamma(a\to e^+ e^-) \approx \frac{1}{8 \pi} \frac{Q_e^2 m_e^2}{f_a^2} m_a,
\end{align}
we obtain the lifetime of axion in the lab frame 
$\gamma c\tau_a \approx 1.65~{\rm cm}$ for $Q_e = 1/2$ and $m_{a}=17$~MeV.

Since the axion $a$ is pretty light, the positron and electron pair (we will refer to them as electrons collectively) in the $a\to e^+ e^-$ decay will be highly collimated with a separation of 
$\theta \sim m_a/E_a \approx 3\times 10^{-4}$ radian at the decay vertex. With the 3.8 Tesla magnetic field in the tracker, the two tracks will separate with the curvature given by $R =|p_T/(q\times B)|=(p_T/{\rm GeV})\times 0.88$ meter.
The separation angle between two tracks at the EM calorimeter cells will be $\theta_{\rm sep} \approx 2 \arcsin(r/(2R))$, where $r$ is the radius for the EM calorimeter cells located in the transverse plane. 
Assuming the electrons have $p_T\sim$ 10 GeV, this angle will be as large as $0.11$ radian and the two tracks can in principle be recognized as electron and positron with the correct charge assignment. However, the separation is still smaller than the cone size of the typical EM cluster $\Delta R = 0.2$ \cite{Aaboud:2019ynx}. 
Therefore, it is likely to be constructed as an EM cluster. 
The displaced vertex of $a$ can be reconstructed, which is displaced to the primary vertex by $\mathcal{O}({\rm cm})$.
This is similar to the signature of a converted photon in SM events, where the photon passing through the material in the detector has been converted into an electron-positron pair.
In that case, one can sum the momentum of $e^+$ and $e^-$ and obtain the momentum
for the converted photon. Extrapolating the direction of the momentum from the vertex, it will go back to the primary vertex~\footnote{ This is different from the case analyzed in Ref.~\cite{CMS-PAS-EXO-14-017}, where the converted photon has a large impact parameter.}. Therefore, the converted photon has zero impact parameter. 
In the case of the axion, since it will be highly boosted, we can assume the electrons in its decay are parallel.
The electron tracks can have  non-zero impact parameters, 
\begin{align}
	d_0 = \sqrt{R^2 + d_a^2}-R \approx \frac{d_a^2}{2 R},
\end{align}
where $d_a$ is the distance between the displaced vertex (where $a$ decay) and primary vertex in the transverse plane.
Since $d_a \sim \mathcal{O}({\rm cm})$, we see $d_0 \sim 10^{-3} ~{\rm cm}$ which is too close to measure.
Therefore, a highly boosted $a$ decaying into $e^+ e^-$ indeed looks similar to the converted photon in SM event.
But there is one critical difference in that the conversion vertex for the signal depends only on the lifetime of $a$,
while in SM events it has to be in the detector material.

Refs.~\cite{Agrawal:2015dbf, Dasgupta:2016wxw, Tsai:2016lfg} have considered the case of a light and highly boosted $ A'$, which decays into electron pairs, $A' \to e^+ e^-$. Such $A'$ can be reconstructed as converted photon, 
as we previously discussed for light $a \to e^+ e^-$ decay.
Ref.~\cite{Tsai:2016lfg} further pointed out that whether it is reconstructed as a converted photon highly depends on the $A'$ decay location. If it decays before the first layer of the pixel tracker (about 3.4 cm away from the central axis of the detector in the barrel region), the electrons (including positrons) will leave hits in the first layer of pixel detector. In this case,  since the conversion occurs outside the material the event cannot be reconstructed as a converted photon. 
In the flowchart for reconstruction of electrons or photons \cite{Aad:2019tso}, such event is also classified as ``ambiguous". Thus, as the electrons cannot pass the isolation criteria, this will be identified as a lepton-jet signature. 
If $A'$ decays after the first layer of the  pixel detector but before the 3rd-to-the -last layer of silicon-strip detectors (SCT), it can be identified as a converted photon. 

In our scenario, the lifetime of $a \to e^+ e^-$ in the lab is about $\sim 1.65$ cm for the benchmark.   
Therefore, it has the probability $1- e^{- 3.4{\rm cm}/(1.65{\rm cm})}\approx 87\%$ to decay before the
first layer of the pixel detector. As a result, it will dominantly look like a lepton-jet event.
There is a $\sim13\%$ probability to decay after the first layer of the pixel detector. From Ref.~\cite{Aad:2019tso},
the two Si tracks (SCT tracks) have an efficiency of 0.35 to be reconstructed as converted photon for  
$E_T^{\rm true} \sim 20$ GeV. Therefore, with two $a$ both reconstructed as converted photons, the probability is as small as 0.002. As a result, we conclude that the signal $h \to a a$ with $a \to e^+ e^-$ will be negligible in
the contribution to the $h \to \gamma \gamma$ branching ratio. It also does not fit to the $h \to e^+ e^-$ searches because
the electrons in the signal is not isolated.

The lepton-jet signature has been generally discussed and also specifically for two lepton-jets from the SM Higgs decay in Refs.~\cite{ArkaniHamed:2008qp, Baumgart:2009tn, Ruderman:2009tj, Cheung:2009su}. The ATLAS and CMS collaboration have searched for lepton jets from a light boson~\cite{Aad:2012kw,Chatrchyan:2012cg, Aad:2014yea, Sirunyan:2018mgs, Aad:2019tua}, where the lepton jets come from the decay topology $h \to \chi_2\chi_2 \to \chi_1 \chi_1 A'A'$, where $A'$ is the dark photon, and $\chi_2$ is an excited fermion DM, which decay promtly to $\chi_1$, and
$\chi_1$ is the DM, recognized as missing energy at the collider. The $A'$ decay is displaced but Refs.~\cite{Aad:2012kw,Chatrchyan:2012cg, Aad:2014yea, Sirunyan:2018mgs, Aad:2019tua} focused on the di-muon channel only. 
The prompt electron-jet signature has been studied in the ATLAS search \cite{Aad:2013yqp} focusing on the topology $h \to X X,~ X \to A' A'$ with $A'\to e^+ e^-$. There, a constraint is set on the electron-jets decay
BR$(h \to {\rm e-jets} ) \lesssim 0.2$ for $m_{A' } \ge 100$ MeV. However, this search does not cover lower massess $\mathcal{O}(10)$ MeV and the topology is different from our model. 
Therefore, we conclude that $h\to a a$ for an axion with mass $\mathcal{O}(10)$ MeV is still viable and provides an interesting electron-jet signature for future searches.

\section{Conclusions}
\label{sec:conclusions}

The strong CP problem is one of the most intriguing aspects of the Standard Model of particle physics.  It can be naturally solved by 
the introduction of an axion field, implying the presence of a new pseudo-scalar particle in the spectrum.  Generically, in order to 
avoid experimental constraints, the axion decay constant is assumed to be much larger than the weak scale, implying a very 
light axion particle.  In this article, we have investigated the suggestion that in spite of these constraints, the axion may be
as heavy as 17 MeV, opening the possibility that this axion particle may lead to an explanation of the di-lepton resonance observed
in nuclear transitions at the Atomki experiment. For this to happen, the axion should couple only to the first generation quark and
leptons, and should have a small mixing with the pion, something that may be achieved for values of the ratio of the PQ charges
of up and down quarks $Q_{u}/Q_{d} = 2$.  

One of the most relevant constraints on this scenario is imposed by the modifications to the anomalous magnetic moment of the electron. In this
work, we demonstrated that the existence of relevant contributions at the two-loop level may lead to a compatibility of the electron
$(g-2)_{e}$ with the recent determinations of the fine structure constraint. This implies a correlation between the electron coupling
to the axion, and hence its charge under PQ, and the coupling to photons at low energies, induced mostly by the mixing with 
the light mesons, $\pi$ and $\eta$.  Moreover, the corresponding range of couplings leads to PQ charges of order one and that
take values between $Q_{e}/Q_{d} = 0.28 \mbox{--} 2$.

Building a realistic model leading to such a heavy axion is a non-trivial task, and demands a more precise relation between the PQ charges in
the lepton and quark sectors. Concentrating on the charged lepton and quark masses, we build a model based on the introduction
of three additional $SU(2)_L$ doublets and singlets, associated with the up and down quarks and the electron sectors, respectively.   
These fields acquire vacuum expectation values partly induced by interactions with the Standard Higgs doublet, with the doublet vevs 
being much smaller than the singlet ones. The axion field is then mostly associated  with a combination of the singlet pseudo-scalar
components, and their vevs  should be of the order of 1 GeV to generate the axion decay constant $f_a = {\cal{O}}$(1GeV). 
This scenario leads naturally to small first generation quark and lepton masses. 
The axion couples to first generation fermions dominantly with suppressed effective PQ charge to 2nd and 3rd generations through the mixing with SM Higgs. Thus, searches for the axion in the exotic decays of heavy mesons are safely evaded.
In order to avoid a large coupling of the SM Higgs to two axions, the Higgs trilinear couplings with the new singlets and doublets cannot be too large, leading naturally to a relatively 
light spectrum of doublet and singlets with masses of the order of several hundreds or tens of GeV, respectively.   

The model we constructed generates the right quark and charged lepton masses and is consistent with the observed CKM mixing.
It also avoids all experimental constraints and leads naturally to a decay mode of the standard model Higgs into two axions, that
due to the large boost and its corresponding lifetime, may be searched for in the electron channel in ways that are not 
yet studied by experiments.

\section{Acknowledgments}
We would like to thank Wolfgang Altmannshofer, Daniele Alves, Stefania Gori, David Morrissey, and Tim Tait for useful discussions and comments.
CW and NM have been partially supported  by the U.S. Department of Energy under contracts No. DEAC02-06CH11357 at
Argonne National Laboratory.  The work of CW at the University of Chicago has been also supported by the DOE grant DE-SC0013642. TRIUMF receives federal funding via a contribution agreement with the National Research Council of Canada.
The work of JL is supported by National Science Foundation of China under Grant No. 12075005
and by Peking University under startup Grant No. 7101502458.
The work of XPW is supported by National Science Foundation of China under Grant No. 12005009.

\bibliographystyle{utphys}
\bibliography{ref}

\end{document}